\documentclass[10pt,aps,pre,twocolumn,amsmath,showpacs,superscriptaddress,nofootinbib]{revtex4-1}
\usepackage{graphicx,txfonts}
\def\ER{Erd\H{o}s-R\'{e}nyi}
\def\T{\tilde}
\def\s{\sigma}
\def\X{\chi}
\def\k{\kappa}

\def\lb{\left}
\def\rb{\right}
\def\nn{\nonumber}

\def\eij{a_{ij}}
\def\eoj{a_{1j}}
\def\edj{a_{2j}}
\def\etj{a_{3j}}

\def\lij{l_{ij}}
\def\loj{l_{1j}}
\def\ldj{l_{2j}}
\def\ltj{l_{3j}}

\def\XXij{\xi_i-\xi_j}

\def\XXderF{(\xi_i-\xi_j)'_u}

\def\rij{r_{ij}}
\def\rji{r_{ji}}

\def\uij{u_{ij}}

\def\Oij{O_{ij}}
\def\Oji{O_{ji}}
\def\Ooj{O_{1j}}
\def\Odj{O_{2j}}
\def\Otj{O_{3j}}

\def\Sij{S_{r_{ij}}}
\def\sgn{\text{sgn}}
\def\xbar{\bar \xi}
\def\Dbar{\bar D}

\begin{document}
\title{Analytical solution for a class of network dynamics with mechanical and financial applications}

\author{P. Krej\v c\'\i}
\affiliation{Institute of Mathematics, Academy of Sciences of the Czech Republic, Prague, Czech Republic}

\author{H. Lamba}
\affiliation{Department of Mathematical Sciences, George Mason University, Fairfax, USA}

\author{S. Melnik}
\affiliation{MACSI, Department of Mathematics \& Statistics, University of Limerick, Ireland}

\author{D. Rachinskii}
\affiliation{Department of Applied Mathematics, University College Cork, Ireland}
\affiliation{Department of Mathematical Sciences, University of Texas at Dallas, Richardson, Texas 75080, USA}


\begin{abstract}

We show that for a certain class of dynamics at the nodes the response of a network of any topology to arbitrary inputs is defined in a simple way by its response to a monotone input. The nodes may have either a discrete or continuous set of states and there is no limit on the complexity of the network. The results provide both an efficient numerical method and the potential for accurate analytic approximation of the dynamics on such networks. As illustrative applications, we introduce a quasistatic mechanical model with objects interacting via frictional forces, and a financial market model with avalanches and critical behavior that are generated by momentum trading strategies.

\end{abstract}
\maketitle

\section{Introduction}
Dynamical processes on networks are used to model a wide variety of phenomena such as the spreading of opinions through a population~\cite{net1}, propagation of infectious diseases~\cite{net2}, neural signaling in the brain~\cite{net3}, and cascading defaults in financial systems~\cite{net4}. Similar dynamical processes on regular lattices are used for modeling phase transitions and critical phenomena in statistical mechanics~\cite{is1}, avalanches and propagation of cracks in earthquake fault systems~\cite{is2}, percolation phenomena~\cite{is3,Schroeder13}, crackling noise~\cite{is4,Schroeder13} and hysteresis in constitutive relationships of various materials~\cite{is5}. The structure of the underlying network may strongly influence the dynamics, the response of the network to variations of the input and parameters, and the critical values of parameters such as the critical temperature of the random field Ising spin-interaction model~\cite{is6}, or the epidemic threshold for disease-spread models~\cite{net5,net6,net7}. Prediction of the response of a network to variations of the input or initial state is thus an important problem, which remains open for many real-world and randomly generated networks (e.g., networks with arbitrary degree distribution)~\cite{n0}.

Nodes of the above networks are often assumed to have a binary response modeled by Heaviside step functions~\cite{watts}. In this paper, we consider networks with a different type of nodes characterized as Prandtl-Ishlinskii (PI) operators.\footnote{The classical Prandtl-Ishlinskii model of plasticity and friction~\cite{pi1, Visintin94} introduced independently by Prandtl (1928) and Ishlinskii (1944) is obtained by the linear superposition of simple hysteresis operators (stops) that model non-interacting fibers with possibly different physical properties. Recently, the model has found new applications in such areas as control of sensors and actuators~\cite{Kuhnen03,Janaideh11}. The celebrated Preisach model used in modeling ferromagnetism~\cite{Mayergoyz91, Radons08}, magnetostriction~\cite{Davino13}, and porous media flow~\cite{Appelby09} also can be considered as a nonlinear generalization of the Prandtl-Ishlinskii model. The PI operator introduced in this paper generalizes the classical Prandtl-Ishlinskii model by including a possibility of discontinuous response that models avalanches. The objective of this paper is twofold: first, to present a new method for solving dynamics on networks (with arbitrary complex topology) and, second, to explore how the standard models of hysteretic phenomena which in most cases assume no interaction between elementary hysteresis operators will be affected by the interaction of these operators.}

We present an almost explicit solution for the input-state-output relationship for networks of PI operators at the nodes. Essentially, we demonstrate that the network of PI nodes is also a PI operator with, possibly, a discontinuous response. This fact sets a limitation on the class of systems that can be modeled by a network of connected PI nodes while simultaneously providing us with an effective tool for mapping the network topology to its dynamics. Two motivating examples, one with a mechanical and one with a financial background, will be considered.

\section{Mechanical example} \label{sec2}
\begin{figure}[t]
\includegraphics[width=0.97\columnwidth]{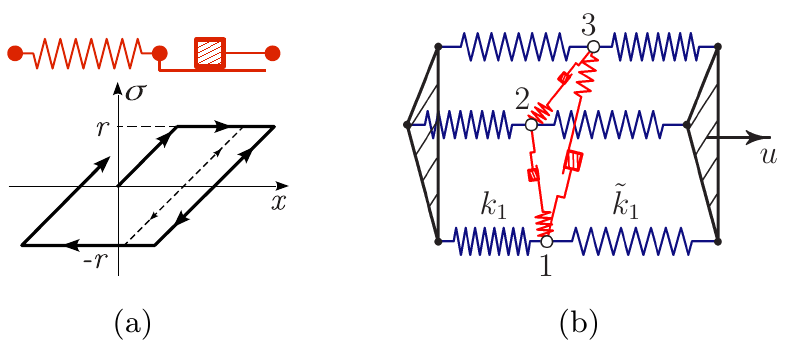}
\caption{(Color online) (a) A mechanical analogy of the stop operator: an ideal spring and an object on a dry surface connected in series. When the spring stress $\sigma$ is within the range $(-r, r)$, variations of the displacement $x$ cause linear changes in $\sigma$ while the object remains stationary on the surface. The spring stress clamps at a value of $\pm r$, whereas the object moves relative to the surface following $x$. (b) A mechanical model with three nodes, each attached to a fixed left plate and a moving right plate by two elastic springs, with interactions modeled by stop operators as in (a).}
\label{fig1}
\end{figure}

In a mechanical context, the PI model describes the hysteretic relationship between strain $x$ and stress $\sigma$ in elasto-plastic materials~\cite{kras}. The simplest example is Prandtl's elastic-perfect plastic element~\cite{pi1}, which combines the restriction $-r\le \sigma\le r$ with the assumption that Hooke's law is obeyed when $|\sigma|<r$. The operator $S_r$ that transforms the input time series $x(t)$ into the output time series $\sigma(t)=S_{r}[x](t)$ of Prandtl's element is called a stop. Figure~\ref{fig1}(a) shows the underlying mechanical model as a cascade connection of a Coulomb friction element and an ideal elastic element, as well as the parallelogram-shaped hysteresis loops in the $(x,\sigma)$ plane. In the Coulomb friction model the force $\sigma$ increases (without causing motion) until it reaches the limit value $\sigma=\pm r$ at which point motion starts and the force remains constant.

\begin{figure}[!b]
\includegraphics[width=0.85\columnwidth]{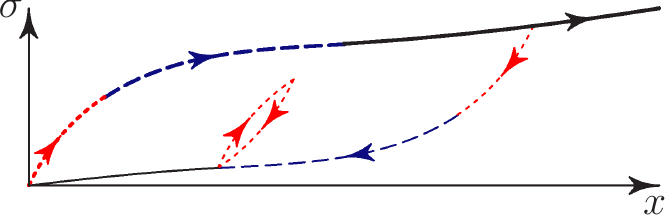}
\caption{(Color online) Loops of the PI operator obtained from the PR curve which is shown by the thick line. Each hysteresis branch (dotted, dashed, and solid curves) is a shifted (or shifted and rotated by 180 degrees) image of the corresponding segment of the PR curve.}
\label{figPI}
\end{figure}

In the general PI model stops with different limits $r$ are superposed so $\sigma(t)=\int_0^\infty S_r[x](t)\,d \mu(r)$, where $\mu$ is some cumulative distribution function. According to this relationship, a new hysteresis loop is initiated in the $(x,\sigma)$ plane each time the input $x$ makes a turning point, see Fig.~\ref{figPI}. Like the Ising and Preisach models~\cite{sethna, mayergoyz}, the PI model has return point memory, which means that the moment the input repeats its past extremum value a hysteresis loop closes and the dynamics proceeds as if there were no such loop~\cite{pi1}. Moreover, the shape of all loops is defined explicitly by the primary response (PR) function {$R(x)= 2\int_0^{x/2} (\mu(\infty)-\mu(r)) dr$}. Namely, for every loop, the arc where the input increases is a shifted initial segment of the graph of the PR function, while the arc of the loop where the input decreases is centrally symmetric to the arc where the input increases, see Fig.~\ref{figPI}. These properties allow one to map an arbitrary piecewise monotone input $x(t)$ to the output $\sigma(t)$ graphically very simply using the PR curve. Equivalently, one can use the sequence of running main extrema $X_k(t)$ of the input $x(t)$ {(see~\cite{mathz})}
\begin{equation}\label{a}
\sigma(t)=\frac{R(2X_1(t))}2+\sum_{k\ge 1} (-1)^k R\bigl(|X_{k+1}(t)-X_{k}(t)|\bigr),
\end{equation}
where we assume zero initial output of each stop $S_r$ and a non-negative input with $x(0)=0$. Here, the running main extrema are defined consecutively as $X_k(t)=\max_{\tau_{k-1}\le \tau \le t} x(\tau)$ for odd $k\ge1$ and $X_k(t)=\min_{\tau_{k-1}\le \tau \le t} x(\tau) $ for even $k\ge 1$, where $\tau_0=0$ and $\tau_k$ is the last moment prior to $t$ when $x(\tau_k)=X_k$.

For any, possibly discontinuous, function $R(x)$ with $R(0)=0$ that has bounded variation, the input-output relationship defined by Eq.~\eqref{a} (equivalently, by Fig.~\ref{figPI}) will be called the PI operator $I_R$ with PR function $R$ and will be denoted $\sigma(t)=I_R[x](t)$. The stop and the PI model are PI operators.

In the PI model the stops do not interact but interactions are necessary for producing more complicated hysteresis loops. Examples of complex hysteretic responses due to interactions include spin-interaction models~\cite{is1,is2}, the moving Preisach hysteresis model~\cite{movingP}, and networks of non-ideal relays \cite{kras}. Such interactions make the models far less tractable and the identification of model parameters extremely difficult. Hence the absence of interactions between the elementary hysteretic components of the model (such as stops or relays) has been considered a necessary simplification in the majority of phenomenological models of hysteresis. However, we will show that networks of PI operators (including systems of interacting stops) are analytically tractable under broad and well-defined assumptions.

We now proceed with an example of a network of interacting stops modeling quasistatic one-dimensional dynamics of a mechanical system that consists of $N$ rigid fibers elongated along the $x$ direction and interacting due to friction between them. The fibers are stretched between two plates; the left plate is fixed, and the right plate is subject to a time dependent quasistatic loading. In Fig.~\ref{fig1}(b), each fiber is represented by a node ($N=3$) attached to two plates by linear springs. The interaction between the nodes is modeled by Maxwell-slip friction elements~\cite{maxwell}. The balance of forces at each node can be written as
\begin{equation}\label{e:MechModel}
-k_i \xi_i + \T k_i(u-\xi_i)+\sum_{j=1,\ldots,N;\ j\ne i} a_{ij} S_{r_{ij}}[\xi_j-\xi_i]=0,
\end{equation}
where $\xi_i$ are displacements of the nodes, the displacement $u$ of the right plate is the time-varying input, $k_i$ and $\T k_i$ are the stiffnesses of the springs attached to the left and right plates respectively, and all the initial displacements and forces are zero. According to the action-reaction principle, the matrix $r_{ij}$ and the adjacency matrix $a_{ij}$, which quantify the strength of the interactions between the nodes via stiction and kinetic friction, are symmetric and non-negative. The system dissipates energy due to friction and the internal energy of the system is $U=\frac12\sum_i (k_i \xi_i^2 +\T k_i (u-\xi_i)^2) + \frac12\sum_i \sum_{j<i} a_{ij} (S_{r_{ij}}[\xi_j-\xi_i])^2$.

\begin{figure}[!ht]
\includegraphics[width=0.97\columnwidth]{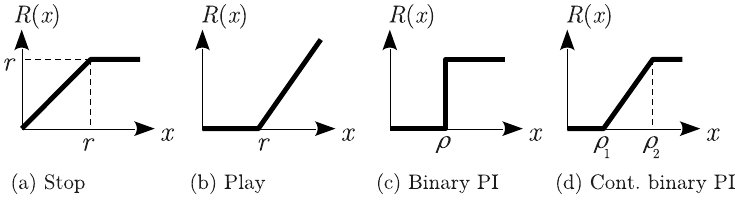}
\caption{PR curves $R(x)$ versus input $x$ for several examples of PI operators: (a) stop, (b) play, (c) binary PI operator, and (d) continuous approximation of a binary PI operator.}
\label{f_PrimResp}
\end{figure}

Our main observation is that if, in response to an increasing input $u$ each distance $|\xi_i-\xi_j|$ corresponding to a nonzero $a_{ij}$ grows monotonically, then the relationship between each displacement $\xi_i$ and the input $u$ is described by a PI operator $I_{R_i}$ for all possible inputs $u(t)$. This fact is rooted in the composition formula~\cite{krej} which ensures that the cascade connection $\sigma=I_{R_1}[I_{R_2}[u]]$ of two PI operators with PR functions $R_1$ and $R_2$, where $R_2$ is monotone, is itself a PI operator $I_{R_1 \circ R_2}$ with the PR function $(R_1 \circ R_2)(u)=R_1(R_2(u))$. Substituting the relations $\xi_i(t)=I_{R_i}[u](t)$ in Eq.~\eqref{e:MechModel}, using the composition formula, and replacing PI operators with their PR functions, we obtain the algebraic system $\T k_i u-(\T k_i + k_i)R_{i}(u)+\sum_{j\ne i} a_{ij} \phi_{r_{ij}}(R_{j}(u)-R_{i}(u))=0$ for the PR functions $R_{i}$ of the PI operators $I_{R_i}$ describing the displacements of the nodes where $\phi_{r}$ is the PR function of the stop $S_r=I_{\phi_r}$, see Fig.~\ref{f_PrimResp}(a). The Browder-Minty property~\cite{minty} of these equations ensures that all the PR functions $R_{i}$ are continuous and increasing. These functions are measurable from the system's response to an increasing input $u$ since $\xi_i(u)=R_{i}(2u)/2$.

\begin{figure}[!b]
  \includegraphics[width=0.97\columnwidth]{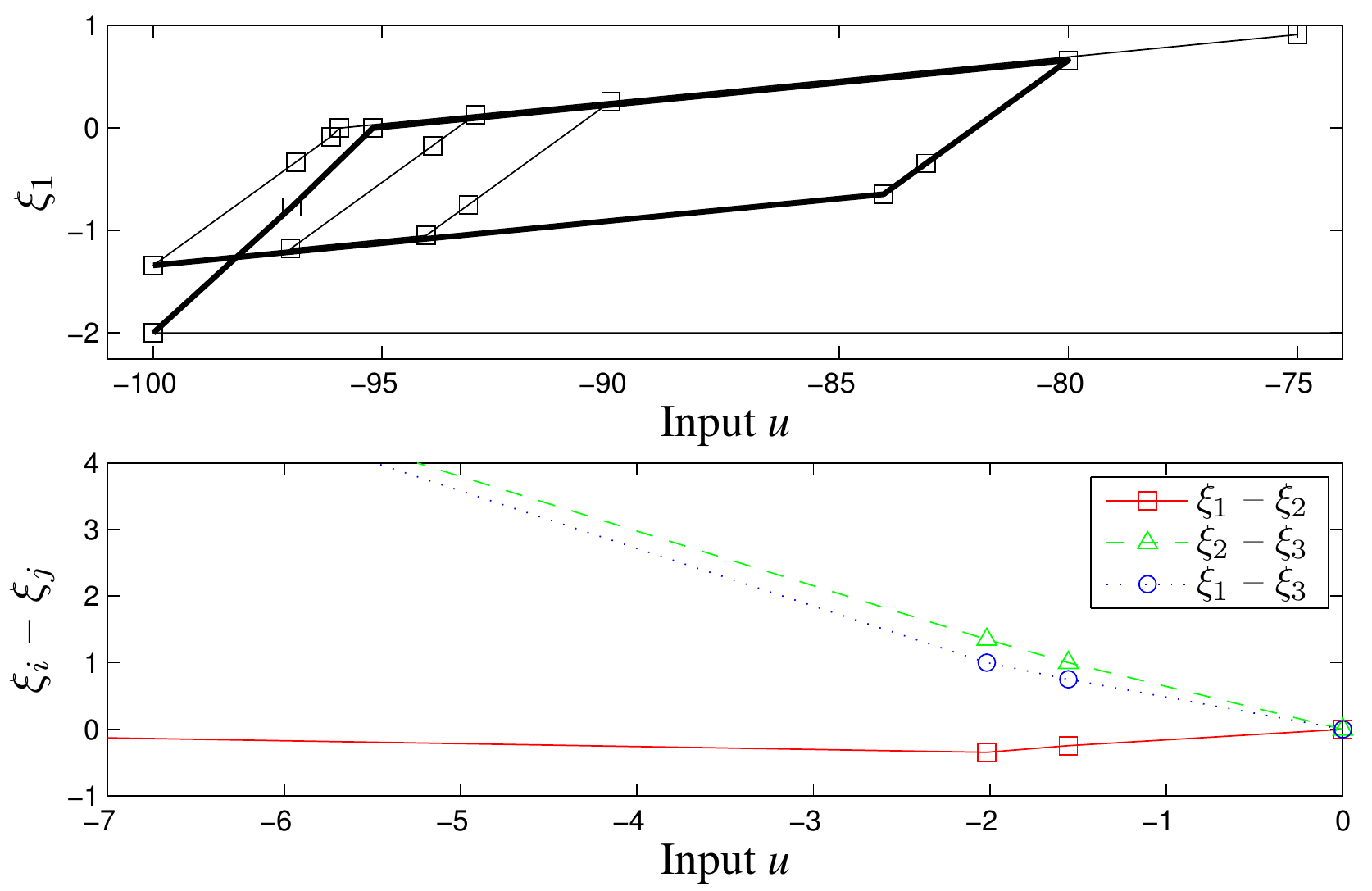}
  \caption{(Color online) An example where $\xi_1-\xi_2$ is nonmonotone for a decreasing input $u$ (the lower panel) and so the relationship between $u$ and $\xi_1$ loses the return point memory property (the non-closed loop shown by the bold line on the upper panel). In this example the system consists of three fibers (nodes) as we show in Fig.~\ref{fig1}(b). Each node interacts with the other two and the forces of interaction between them are $1$ (i.e., all $\eij=1$), and all stop operators have the same $\rij=1$. The left springs' stiffness parameters are $k_1=1, k_2=10, k_3=1$ and the right springs' stiffness parameters are $\tilde k_1=0, \tilde k_2=1,\tilde k_3=10$. Initially all displacements are zero. The values of $u$ at which stop operators saturate or desaturate (see Appendix~\ref{appA}, Table~\ref{table1}) are indicated by symbols.}
  \label{f:mech_non_PI}
\end{figure}

Monotonicity of the relative displacements $\xi_i-\xi_j$ with increasing $u$ is a substantial condition for ensuring the PI relationships $\xi_i(t)=I_{R_i}[u](t)$ between the displacements of nodes and plates in system~\eqref{e:MechModel} for arbitrary inputs $u(t)$. Even in a system of three nodes the differences $\xi_i-\xi_j$ can be nonmonotone in $u$, in which case the relationship between $\xi_i$ and $u$ loses the return point memory property and becomes more complex. Figure~\ref{f:mech_non_PI} presents an example of such behavior. Here the relative displacement $\xi_1-\xi_2$ between the nodes $1$ and $2$ changes nonmonotonically when the input increases (decreases); see the lower panel. As a result, the relationship between the input $u$ and displacement $\xi_1$ time series is not of a PI form: When the input $u$ changes, for example, from $-100$ to $-80$ and back to $-100$, the hysteresis loop does not close as shown by the bold line in the upper panel (see Appendix \ref{appA} for details).

\begin{figure}[!t]
  \includegraphics[width=0.97\columnwidth]{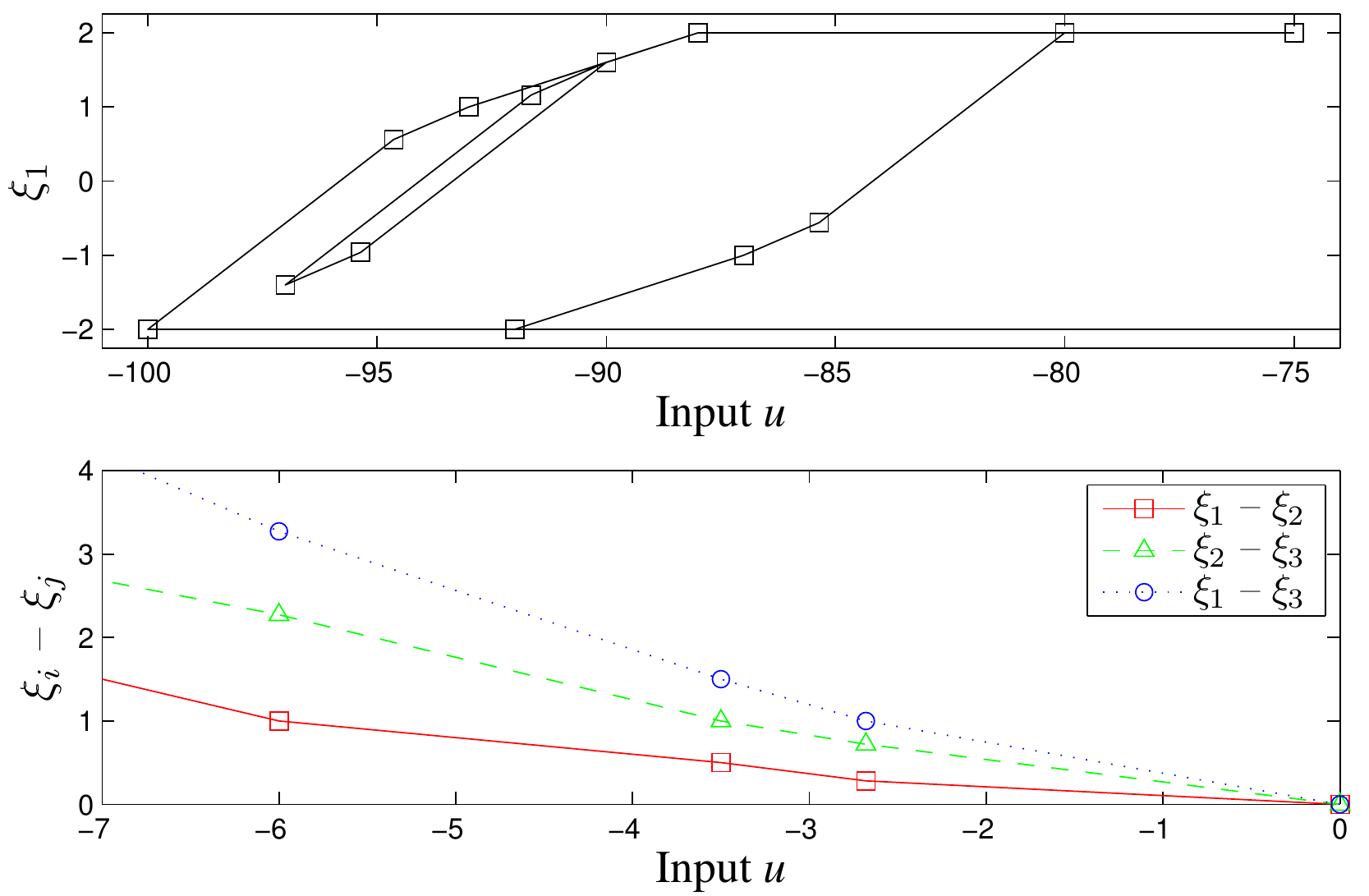}
  \caption{(Color online) An example where $\xi_i-\xi_j$ are monotone in $u$, hence $\xi_i(t)=I_{R_i}[u](t)$. The network structure, parameters and the variation of $u$ are the same as in Fig.~\ref{f:mech_non_PI} except that $k_1=k_2=k_3=1$. The upper panel shows variations of the position $\xi_1$ of the first node in response to the input $u$ which starts at $0$ and varies monotonically between the following turning points: $\{0, -100, -80, -100, -90, -97, -75\}$. The values of $u$ at which stop operators saturate or desaturate (see Appendix~\ref{appA}) are indicated by symbols. Plots of $\xi_2$, $\xi_3$ against $u$ (not shown), as well as plots of any weighted sum of $\xi_i$, also demonstrate symmetric loops. The lower panel shows the monotonic growth of the displacements $|\xi_i-\xi_j|$ for a decreasing input $u$ starting at $0$.}
  \label{f:mech_PI}
\end{figure}

However, if all the friction forces are relatively small compared to the forces of the springs, then the distances $\xi_i-\xi_j$ are monotone and $\xi_i(t)=I_{R_i}[u](t)$. For example, Fig.~\ref{f:mech_PI} presents a system of three interacting fibers (nodes) where all three relative displacements $|\xi_i-\xi_j|$ grow monotonically in response to an increasing (decreasing) input $u$ (see the lower panel). Hence, the position of each node $\xi_i$ is related to the displacement of the right plate $u$ by a PI operator $\xi_i(t)=I_{R_i}[u](t)$. Indeed, all the hysteresis loops in Fig.~\ref{f:mech_PI} (see the upper panel) are closed and centrally symmetric, which is the characteristic property of PI operators.

In other words, weak interactions merely correspond to parameter changes in the Prandtl-Ishlinskii model and so cannot induce any extra complexity in the model response. This scenario provides a plausible explanation for why the simplified phenomenology underlying the Prandtl-Ishlinskii model gives good approximations across multiple applications~\cite{pi1,Visintin94,Kuhnen03,Janaideh11}. However, stronger interactions generate more complex responses as in the example in Fig.~\ref{f:mech_non_PI} which exhibits the phenomenon of ratcheting (accumulating nonclosed hysteresis loops) which cannot occur in any Prandtl-Ishlinskii model. Note that that standard models of ratcheting used, for example, in the study of fatigue and damage (see, e.g., Section 5.4.4 of~\cite{Lemaitre90}), combine the Prandtl-Ishlinskii model with an additional nonlinearity.

An algorithm for the simulation of systems such as \eqref{e:MechModel} is presented in Appendix~\ref{appA}.

\section{Financial example}
In this section, we use PI networks (with discontinuous PR functions) to model momentum-based trading strategies within a financial market. We start by describing the simplest version of the model in which traders sell (buy) when the ratio of the price to a running maximum (minimum) of the price hits certain threshold values. This wholly price-based strategy is a plausible proxy for an important subset of real-world traders --- so-called {\em momentum traders},\footnote{{\em Fundamentalist traders}, on the other hand, trade based on calculations of whether a stock is over- or undervalued according to some model of the fair or correct price.} who either (a) believe that the recent price history is signaling an upcoming change or reversals in market ``sentiment''~\cite{drawdown,topia} or (b) have been on the wrong side of the recent price history and feel enough pressure to have to switch their position \cite{harbir2}. Momentum traders tend to act as a source of positive feedback that exaggerates recent price moves and can induce, in a plausible manner, both the long-term mispricings and sudden reversals that are characteristic of financial systems.

We then generalize the model by supposing that the market participants also have a network structure and each agent now reacts not only to the price but to the states of their network neighbors. Once the effect of agents changing investment positions is allowed to feed back into the price the network model makes full use of the results outlined in Sec.~\ref{sec2}.

\subsection{Momentum trading strategies as PI operators}
We consider $N$ traders with the state $\chi_i$ of trader $i$ being either $1$ or $-1$. The ``long'' state $\chi_i=1$ indicates that the $i$-th trader owns the asset and the ``short'' state $\chi_i=-1$ means the trader does not own the asset.

Other traders, not modeled directly, play two important roles. First, many operate on short time scales, comparable with the arrival of new exogenous information, and translate this information into price changes. This allows us to consider the system as being slowly driven through metastable states. Second, they provide a pool of potential trading partners so buyers and sellers among the $N$ traders do not need to be matched (as occurs in kinetic theory models of financial systems).

The following drawup-drawdown rule~\cite{drawdown} for the $N$ traders mimics strategies that try to identify a nascent trend and are used in actual trading algorithms\footnote{The strategy described below, or minor variations of it, are implementable on some trading platforms by placing a {\em trailing stop} order.} (see, e.g., Ref.~\cite{topia}).

After switching to the long state $\chi_i=1$ (purchasing the asset) at time $\tau$, the $i$-th trader tracks the asset price $p(t)$ and the running maximum $\max_{\tau\le s\le t} p(s)$ since time $\tau$. The trader switches back to the short state $\chi_i=-1$ at the first time $\theta>\tau$ when the inequality $p(t)/{\max_{\tau\le s\le t} p(s)} \leq \alpha^-_i$ is satisfied for some threshold value $\alpha^-_i\in(0,1)$. For example, if $\alpha_i^-=0.9$, then the trader sells at the moment when the price drops from its peak value by 10\%. Using the log-price $r(t)=\ln( p(t)/p(0))$ gives the selling condition $ \theta = \min \{t>\tau: r(t)-\max_{\tau\leq s \leq t}r(s) \leq \ln \alpha^-_i\}$. (Without loss of generality we use natural logarithms in this paper.) This trader then adopts a similar strategy for deciding when to buy again. The trader tracks the ratio $p(t)/{\min_{\theta\le s\le t} p(s)}$ and switches to the state $\chi_i=1$ when it exceeds a value $\alpha_i^+>1$.

Following Ref.~\cite{harbir2}, the aggregated quantity $\sigma=\sum_{i=1}^{N}\mu_{i}\chi_{i}$ represents the overall sentiment of the market where the weights $\mu_{i}>0$ are a measure of the market impact of each trader.

To use the results of Sec.~\ref{sec2} we must make the mild assumption that $\ln \alpha^+_i=-\ln \alpha^-_i:=\rho_i$ for each trader. Then the relationship between $r(t)$ and the state $\chi_i(t)$ of each trader is defined by the binary PI operator $\chi_i(t)=I_{H_i}[r](t)$ whose PR function is the shift $H_i(r)=H(r-\rho_i)$ of the step function $H(r)$ [see Fig.~\ref{f_PrimResp}(c)]. Moreover, the sentiment is related to the log-price by the PI operator $\sigma(t)=I_{R}[r](t)$ with the PR function $R(r)=\sum_{i=1}^N \mu_i H_i(r).$

\begin{figure}[!hb]
  \includegraphics[width=0.99\columnwidth]{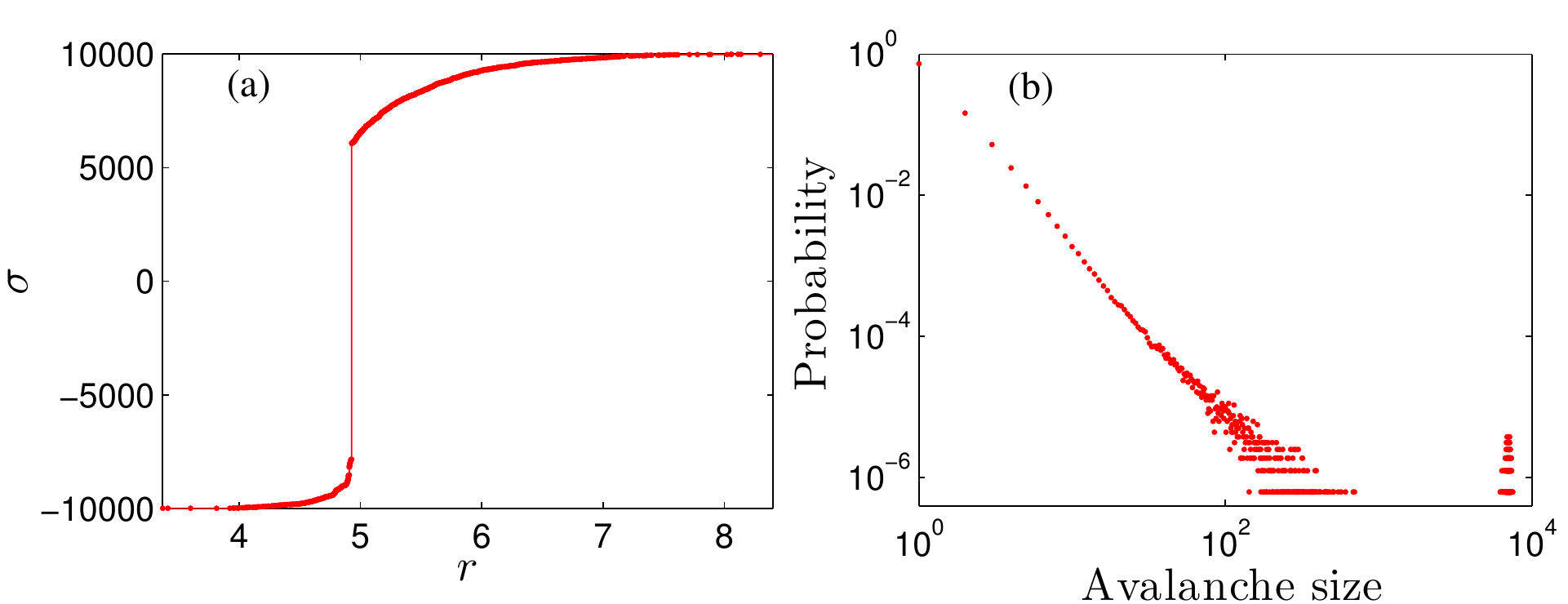}
  \caption{(Color online) (a) PR curve of a network~\eqref{c} of binary PI operators whose PR curves are as in Fig.~\ref{f_PrimResp}(c) (momentum traders). To define the adjacency matrix $a_{ij}$ we use, as an example, an undirected unweighted \ER~network (i.e., a graph in which each pair of nodes is connected by an edge with equal, independent probability) of $N=10^4$ nodes with mean degree $5$. Threshold values $\rho_i$ for the nodes are taken from the normal distribution with mean $7$ and variance $1$. Other parameters are $\mu_i=1$ and $b_i = 1$ for all $i$. We start with $r=0$ and all nodes in state $-1$; we then increase $r$ until all nodes reach state $1$. (b) Size distribution of avalanches exhibited by the same system. The statistics is calculated from 1000 realizations of random networks and $\rho_i$. The spike in the distribution for large avalanche sizes corresponds to the large jump in (a).}
  \label{f:fin_model}
\end{figure}

So far each agent's PI operator reacts to the same input, namely the log-price $r(t)$. We now introduce coupling between the traders by replacing the log-price $r$ in the trading strategy of the $i$-th trader with the aggregated quantity $\xi_i=\sum_{j=1}^N a_{ij} \chi_j + b_i r$. This leads to the network model
\begin{equation}\label{c}
\chi_i(t)=I_{H_i}\Big[\sum_{j=1}^N a_{ij} \chi_j(t) + b_i r(t)\Big]; \ \sigma=\sum_{j=1}^N \mu_{j} \chi_j,
\end{equation}
where $b_i,\mu_i\ge 0$. The coefficients $a_{ij}\ge0$ measure the (attracting) influence of the $j$-th trader upon the decision making of the $i$-th trader. Using the composition formula for PI operators (as in the above mechanical example), the solution of model~\eqref{c} takes the form of the PI operator relationship $\chi_i(t)=I_{\hat H_i}[r](t)$ between the state of each trader and the log-price $r$, where the set of thresholds of the step response functions $\hat H_i$ is a subset of the set of thresholds $\rho_i$ of the functions $H_i$. The composition formula for PI operators with continuous PR functions~\cite{krej} requires justification when applied to~\eqref{c} with discontinuous $H_i$ but can be derived using Kurzweil integral theory~\cite{kurzweil}. The PR curve $R(r)=\sum_{i=1}^N \mu_i \hat H_i(r)$ of the PI relationship $\sigma(t)=I_R[r](t)$ between the log-price and the sentiment can be obtained by testing~\eqref{c} with an increasing input $r(t)$ (see Fig.~\ref{f:fin_model}(a)) or by solving the algebraic system
\begin{equation}\label{d}
\hat H_i(r)=H\Big[\sum_{j=1}^N a_{ij} \hat H_j(r) + b_i r-\rho_i\Big]
\end{equation}
derived from~\eqref{c}. A large jump in the PR curve in Fig.~\ref{f:fin_model}(a) corresponds to an avalanche: A change in the state of one node causes other nodes to change their states (via network connections), triggering a cascade.

If we replace the binary PI operator $\chi_i(t)=I_{H_i}[\xi_i](t)$ at the nodes of model~\eqref{c} by the simple input-output relationship $\chi_i(t)=H(\xi_i(t)-\rho_i)$ (a memoryless ideal switch), the response of the network to increasing inputs, i.e., the PR function, remains the same. Hence, the many results describing PR functions of networks of Heaviside switches (such as the statistics of avalanches and critical parameters, see, e.g., Ref.~\cite{is1}) are equally valid for PI networks~\eqref{c}; see Fig.~\ref{f:fin_model}(b). The equation $\sigma(t)=I_R[r](t)$ then explicitly describes the response of the PI network to arbitrary inputs in terms of its PR function $R$, while Eq.~\eqref{d} links the network topology (in terms of its adjacency matrix) with the PR function $R=\sum_{i=1}^N \mu_i \hat H_i$. In particular, the network of binary PI nodes can be set to produce the same response to increasing inputs as any given Ising spin model. However, the response of the Ising model to nonmonotone inputs is more complicated than that of the PI network.

We now compute an example of the network model~\eqref{c} for interacting momentum traders, see Fig.~\ref{f:fin_model} for the parameters of the network. Figure~\ref{f:fin_model}(a) presents the PR curve of the PI relationship $\sigma(t)=I_R[r](t)$ between the logarithmic asset price and the market sentiment (a solution of the model). This PR curve has been obtained simply by testing system~\eqref{c} with an increasing input $r(t)$. The histogram in Fig.~\ref{f:fin_model}(b) shows statistics of avalanche sizes for the PR curve calculated from 1000 realizations of random networks and node thresholds. A large jump in the PR curve corresponds to a big avalanche involving many nodes.

We stress that (large) jumps of the network PR curve $R$ are due to avalanches (caused by interactions between nodes) rather than the discontinuity of the response function $H_i$ at the nodes. A similar discontinuous PR curve $R$ can be generated by a network of the PI nodes with continuous states, where each node has the continuous PR curve shown in Fig.~\ref{f_PrimResp}(d) (PI models of investment (supply) strategies with a continuous PR curve, such as the one shown in Fig.~\ref{f_PrimResp}(b), have been proposed in the economics literature~\cite{goecke}). The counterpart of Eq.~\eqref{d} for a network model with such nodes can result in a PI operator with a discontinuous response caused by avalanches.

It is worth noting that the PR function of the stop operator shown in Fig.~\ref{f_PrimResp}(a) generates clockwise hysteresis loops. This is in contrast to the counterclockwise hysteresis loops produced by the play operator whose PR function is shown in Fig.~\ref{f_PrimResp}(b). PI operators of momentum traders [Fig.~\ref{f_PrimResp}(c)] can generate loops with either orientation.

\subsection{Pricing models}
We can now feed changes in the overall sentiment back into the price to generate asset pricing models. We start with a simple mean-field feedback case where the following simplifying assumptions allow us to compute analytical solutions and describe how the transition from continuous to discontinuous PR curves dramatically changes the market dynamics.
\begin{figure*}[!ht]
\includegraphics[width=1.6\columnwidth]{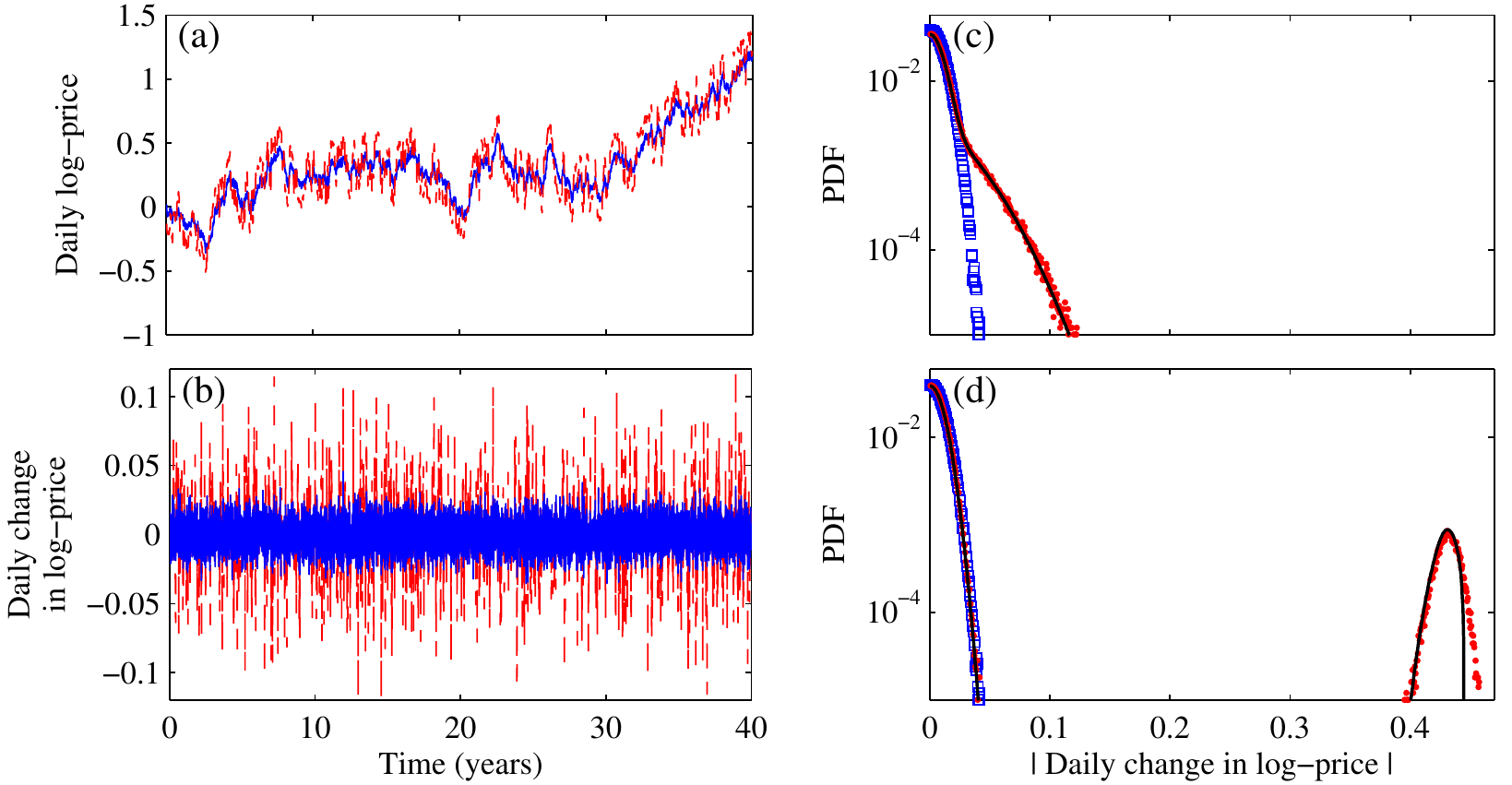}
\caption{(Color online) (a) Time series of the log-price $r^*(t)$ (dashed red line) and the exogenous Brownian information stream $r(t)$ (solid blue line). (b) Daily increments of the log-price $r^*(t)$ (dashed red line) and $r(t)$ (solid blue line). Plots (a) and (b) were obtained for $N=10\,000$ traders with thresholds uniformly distributed over the interval $[c,a] = [0.05,0.45]$ for $\kappa=0.15$. (c) Histogram of the daily log-price increments (red dots) and the exogenous Brownian information stream (blue squares) obtained from 50 simulations with the same parameters as in (b). The black curve is the analytic approximation for $r^*(t)$ (see Appendix~\ref{appC}). (d) Same as (c) but for $\kappa=0.21$, which is slightly above the critical $\kappa_c=0.2$.}
\label{f:pricing_model}
\end{figure*}

It is reasonable to reinterpret $r(t)$ in the definition of $\xi_i$ as being an exogenous Brownian information stream rather than the log-price. The log-price, now denoted $r^*(t)$, is assumed to be modified by the sentiment in a proportional way leading to $r^*(t)=r(t)+\kappa\sigma(t)$, where the parameter $\kappa >0$ quantifies the effect of momentum traders on the price (if, say, more momentum traders enter the market, then $\kappa$ will increase). We choose $\mu_i=1/N$, $a_{ij} = \kappa/N$, and $b_i=1$ so $\chi_i(t) = I_{H_i}[r^*](t)$ and, as before, the traders react solely to the price. Finally, the thresholds $\rho_i$ are chosen uniformly from an interval $[c,a]$. Plausible ranges of the parameters $a$ and $c$ can be estimated as follows. A momentum trader reacting to price changes on the order of, say, $1\%$ would trade too frequently, incurring significant transaction costs, with most of the trading being driven by random fluctuations rather than actual changes in the price trend. Conversely, thresholds of the order of 50\% would result in very infrequent trading that misses many moderately sized trends. The parameter $\kappa$ can be estimated by considering the total influence of momentum traders on the asset price. A reasonable estimate of the difference in price between a market with maximum positive sentiment ($\sigma = 1$) and negative sentiment ($\sigma = -1$) is 20--50\% ceteris paribus (although it may go much higher during an asset bubble as new speculators enter the market: During such an event the distribution of threshold values may also move lower as traders' investing time horizons shorten). The values $[a,c]=[0.05,0.45]$ that have been used in computations for $N = 10\,000$ agents presented in Fig.~\ref{f:pricing_model} are consistent with these estimates.

Explicit calculations are possible in the continuum limit $N\rightarrow \infty$ (the details are available in Appendix~\ref{appC}). The PR curve $R$ of the PI operator $\sigma=I_R[r]$ that relates the Brownian input $r$ to the log-price $r^*=r+\kappa\sigma$ becomes a step function at the critical value $\kappa_c = ({a-c})/{2}$.

The supercritical case $\kappa > \kappa_c$ exhibits extreme jumps between $\sigma = \pm 1$ when all the traders change their state simultaneously [see Fig.~\ref{f:pricing_model}(d)] resulting in a bimodal price change distribution. However, in reality, these systemwide avalanches are unlikely to occur as some of the modeling assumptions will break down. In particular, the market will no longer function with sufficient liquidity (counterparties to a desired transaction may not be available) and the full impact of the avalanche will be spread out over time. A more detailed discussion of such illiquid markets in a related agent-based model can be found in Ref.~\cite{harbir2}.

The subcritical case $\kappa < \kappa_c$ is more relevant to normal market conditions and also more subtle. Here the continuous PR curve of the operator $\sigma=I_R[r]$ has the shape shown in Fig.~\ref{f_PrimResp}(d). The dynamics can be reformulated as a random walk of a particle on a closed rectangular domain with motion along the right (left) boundary corresponding to increasing (decreasing) $\sigma$ and motion on the interior and upper and lower boundaries corresponding to constant $\sigma$ (see Fig.~\ref{rectangle} in Appendix~\ref{appC}). For a fixed $\kappa < \kappa_c$ this model provides an analytic approximation (see Appendix~\ref{appC}) to the distribution of log-price changes over a given time interval such as can be seen in Fig.~\ref{f:pricing_model}(c). The tails of these distributions in actual markets are often claimed to be power laws~\cite{powerlaw} but here they are in fact close to a sum of different Gaussian and error functions.\footnote{A critique of the naive use of linear regression to claim evidence of power laws can be found in Ref.~\cite{clauset}.} For completeness of the mathematical analysis we note that as $\kappa$ approaches $\kappa_c$, the distribution becomes bimodal as in Fig.~\ref{f:pricing_model}(d), where the smaller mode corresponding to large changes of the price separates from the main Gaussian mode.

The existence of a critical value together with the possibility of $\kappa$ varying in time suggests a mechanism for extreme market volatility and the associated bubbles and crashes and fat tails. As a particular asset class receives increased attention or is perceived to be undergoing some fundamental positive change, the price will rise and attract more momentum traders and short-term speculators. This will cause $\kappa$ to increase through the critical value and the system to evolve with $\sigma$ at or close to $+1$ until changes in the process $r^*(t)$ trigger the drawdown process and a systemwide downward cascade.

It is not our aim here to match the fat tails generated by the simple model above with the approximate power laws measured in real, highly complex, financial markets. Rather, we have demonstrated theoretically a plausible mechanism for generating fat tails. The model also predicts that as the proportion of traders who use such a strategy increases, the system will pass through a critical point beyond which a systemic market failure is inevitable. We believe that this model, due to its simplicity and theoretical tractability, complements other heterogeneous agent-based models (see Ref.~\cite{LeBaron00} for examples) that also generate cascades and fat tails but rely solely on numerical simulations.

Finally, we examine and compare some PR curves for a scale-free network model. We also show that the use of the theoretical results from Sec.~\ref{sec2}, together with a numerically computed PR curve, can achieve significant computational savings. We create an undirected unweighted network of $N=10\,000$ nodes (agents) by taking node degrees from the truncated power-law distribution,
\begin{equation}
P_k = \lb\{\begin{array}{rl} \beta k^{-2.5}, & 3\le k\le 50 \\
0, & \text{otherwise} \end{array}\rb.
\label{Pkpowerlaw}
\end{equation}
(with the normalization constant $\beta$ such that $\sum_k P_k = 1$), and then randomly connecting pairs of nodes to obtain the network. Let $a_{ij}$ be the network adjacency matrix. We assign a threshold to each agent from the Gaussian distribution with mean $(a+c)/2$ and variance $1/20$, but we only take values between $c$ and $a$ from this distribution. All the agents are assigned the same weight $\mu_i = 1/N$, see \eqref{c}.

The input of the $i$-th agent is given by
\begin{align}
\xi_i(t) = r(t) + \k \s(t) + \T\k S_i(t),
\label{input_Ii}
\end{align}
where $r(t)$ is the external Brownian input to the system, $\s(t) = \sum_j \mu_j \X_j(t)$ is the sentiment of the market, and $S_i(t) = \sum_j a_{ij} \mu_j \chi_j(t)/\sum_j a_{ij} \mu_j$ is the peer pressure for agent $i$. We define the log-price of an asset at time $t$ as $r^*(t) = r(t) + 0.12\s(t)$, which means that when $\k=0.12$ and $\T\k=0$, the agents make their decisions based solely on the price. When $\T\k>0$, the agents additionally take into account the states of their network neighbors so by varying $\k$ and $\T\k$ we can change weights of the components involved in agents' decision making.

\begin{figure}[!ht]
\includegraphics[width=0.95\columnwidth]{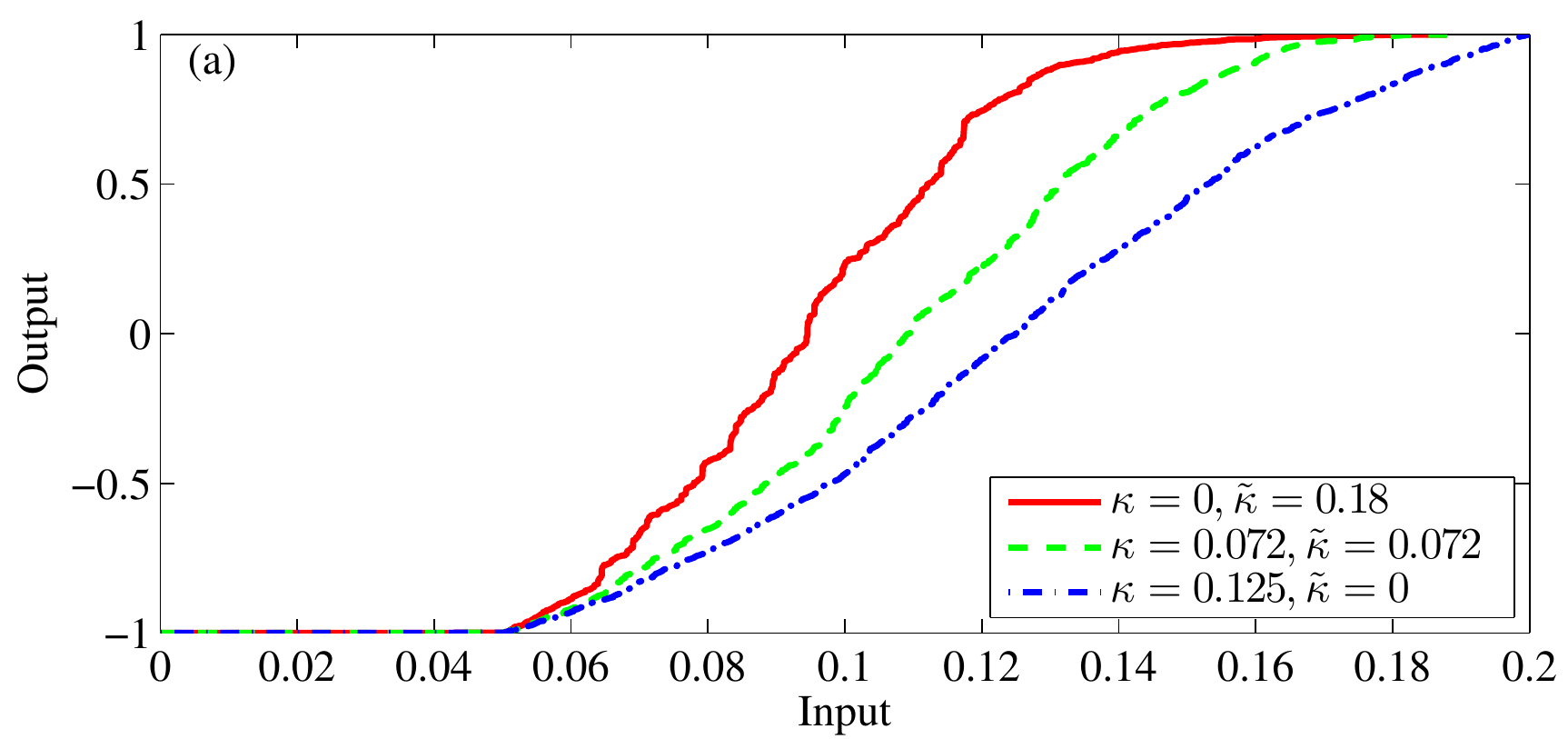} \vskip 2mm
\includegraphics[width=0.97\columnwidth]{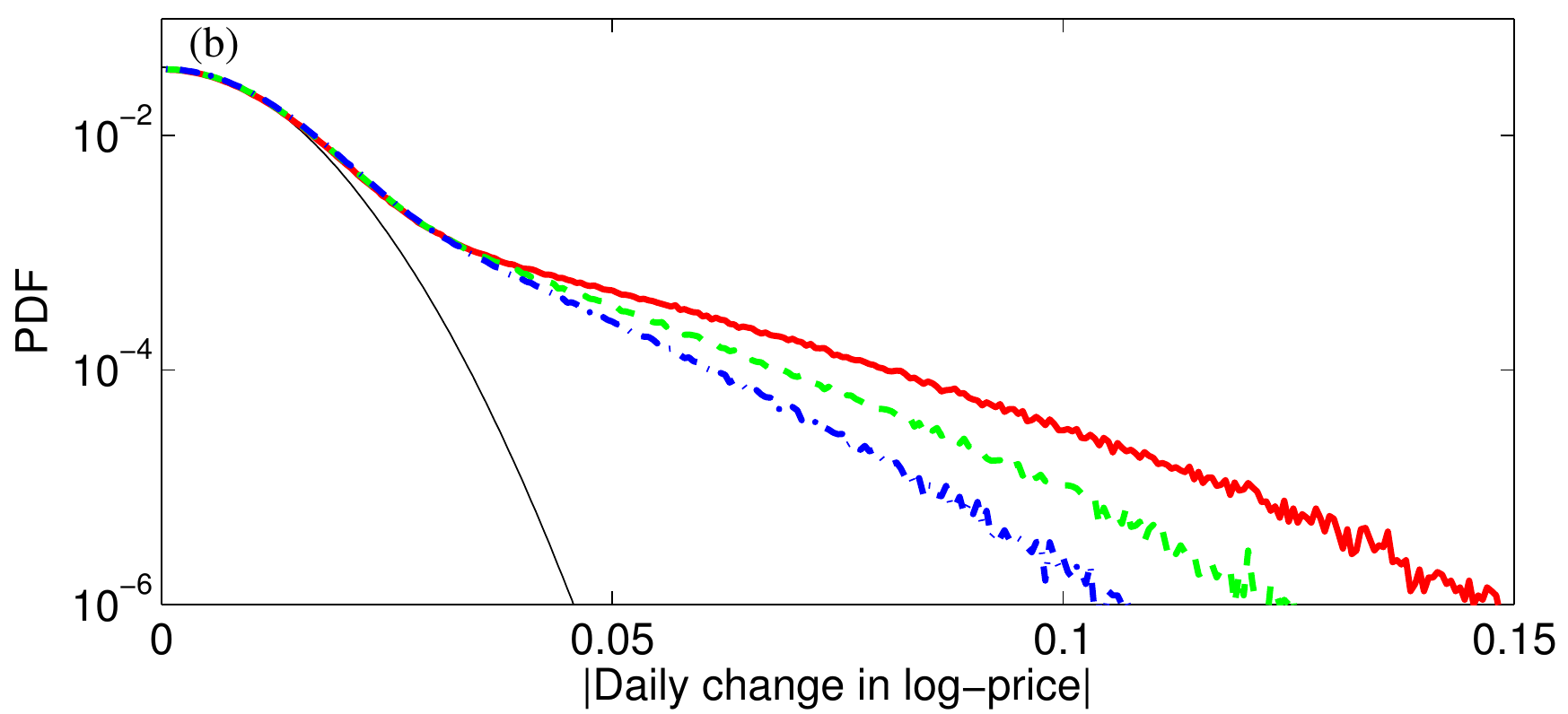}
\caption{(Color online) (a) PR functions for a random network of $10\,000$ agents with degree distribution~\eqref{Pkpowerlaw} for three different pairs of values of $\k$ and $\T\k$ (we chose the values such that all three PR curves reach the saturation value 1 at the input value $0.2$). (b) Histograms of the daily log-price increments $r(t) + 0.12 \s(t)$ (thick curves) obtained from 1000 simulations using the corresponding PR curves in (a). The thin curve shows the distribution of the absolute values of the increments for the exogenous Brownian information stream $r(t)$. The values of other parameters are the same as in Fig.~\ref{f:pricing_model} unless specified otherwise in the text.}
\label{f:PR_Hist}
\end{figure}

Figure~\ref{f:PR_Hist}(a) presents PR functions for a networked system with three different pairs of values of $\k$ and $\T\k$. In order to obtain the PR curves, we start with all agents in state $-1$ and gradually increase the external input $r$ from 0 until all agents are in state $+1$. For each increment of $r$, we let the system reach its stationary state (recall that switching of some agents may increase the input of other agents above their threshold and cause them to switch as well). Once the stationary state is reached, we record the value of $\s=\hat\sigma_i$ and the corresponding value of $r=\hat \rho_i$ (we record these values only if there were any switches). Once all agents switched to $+1$, the set of recorded pairs of $r$ and $\s$ gives us the piecewise constant PR curve $R$.

The operator $\sigma(t)=I_R[r](t)$ that maps the time series of the Brownian information stream to the time series of the market sentiment for the network model can now be understood, and implemented numerically using the above PR curve $R$, as an equivalent system of {\em independent effective} agents $\hat\chi_i$. The number of such agents is equal to the number of discontinuity points in $R$ (which is generally smaller than the number of original agents). Thresholds of independent effective agents are given by the discontinuity points $\hat\rho_i$ of the PR function, while the weight of the $i$-th agent is equal to half the change in the value of the PR function at the $i$-th discontinuity point, $\hat \mu_i=(\hat\sigma_i-\hat\sigma_{i-1})/2$ (that is, $\hat \mu_i$ is the sum of the weights of all the agents in the network that switch collectively as the input increases through the value $\hat \rho_i$). All effective agents are independent of each other, i.e., the input of each effective agent is just the Brownian information stream $r(t)$ [cf. Eq.~\eqref{input_Ii}]. When we replace all agents of the original networked model with the effective agents, the system $\sigma(t)=\sum_i \hat\mu_i \hat\chi_i(t)$ that we obtain will be equivalent to the original networked system (both systems produce the same output $\sigma$ in response to any variation of the input $r$). In other words, we no longer need to consider the network structure because its effect is embedded in the thresholds and weights of the effective agents. This gives us a substantial computational advantage: Not only the number of agents is reduced, but there is no need for computationally expensive calculation of peer pressure, and since the system no longer exhibits cascades of activations it immediately reaches a stationary state for each value of $r$.

Figure~\ref{f:PR_Hist}(b) presents histograms of the daily log-price increments for the network model; they correspond to the PR curves shown in Fig.~\ref{f:PR_Hist}(a). We define log-price as $r(t) + 0.12 \s(t)$ and run 1000 simulations [here we calculate the increments using the system of independent effective agents and not the original system of interacting agents as in Fig.~\ref{f:pricing_model}(c)]. In this example, the fattest tail of the log-price returns distribution is achieved when the pressure of network neighbors has the strongest effect on the decision making of the agents (the largest $\tilde\kappa$). The least-fat tail occurs when the network structure is absent and agents react solely to the price.

\section{Conclusions}
To summarize, we have considered input-driven dynamics on networks with PI operators at the nodes. Examples of such nodes are provided by models of plasticity and friction and some common trading strategies. We have shown that no matter how complex the network, its response to arbitrary variations of the input is described by an effective PI operator and hence can be deduced in a simple and explicit way from the network's response to a monotonically increasing input. Using these results we have shown that one-dimensional models of friction and plasticity with interacting elastic and dry friction elements can be reduced, in case of not-too-strong coupling, to the standard PI model without interactions. We have also derived the analytical form of the fat-tailed price returns induced by momentum-based trading in a financial market. Extending the analysis to allow for the varying influence of momentum traders (the parameter $\kappa$) may yield new insights into the approximate power-law scalings claimed for actual markets. Finally, the numerical method used for our simulations provides a computationally efficient alternative for solving the dynamics on arbitrarily complex networks of PI operators and with arbitrary inputs.

\section*{Acknowledgments}
We thank A. Amann, M. Dimian, and B. Hanzon for useful discussions. This work was funded in part by GA\v CR Grant No. P201/10/2315 and RVO: 67985840 (P.K.), the Irish Research Council (New Foundations grant to S.M.) cofunded by Marie Curie Actions under FP7 (INSPIRE fellowship, S.M.), and Science Foundation Ireland (Grant No. 11/PI/1026, S.M.). D.R. acknowledges the support of NSF through grant DMS-1413223.

\appendix
\section{Simulation of the mechanical model} \label{appA}
In this section we consider in more detail the mechanical model schematically illustrated in Fig.~\ref{fig1}(b) and described by Eq.~\eqref{e:MechModel}. This model can be used to represent a bunch of one-dimensional rigid fibers [shown as nodes in Fig.~\ref{fig1}(b)] elongated along the horizontal axis, whose left and right ends are attached (by springs) respectively to the left and the right plates. The displacement of fiber $i$ relative to the left plate is $\xi_i$. We assume perfect elastic interactions between each fiber $i$ and the left (and the right) plate with coefficients $k_i$ (and $\T k_i$ correspondingly). Furthermore, we assume that each fiber is in contact with some other fibers along its length and there is Maxwell friction when they move with respect to one another. We model the friction force acting on the $i$-th fiber due to its relative displacement with respect to the $j$-th fiber by $\eij \Sij [\xi_j - \xi_i]$, where fiber interaction strengths $\eij $ are non-negative and $\Sij$ denotes the stop operator of half-width $\rij \ge 0$ [see Fig.~\ref{fig1}(a)] with input $\xi_j - \xi_i$. Initially, all forces and displacements in the system are 0. The time-varying input of the system is the displacement $u$ of the right plate relative to its initial position [see Fig.~\ref{fig1}(b)]; the left plate does not move. All the motions are quasistatic.

Equation~\eqref{e:MechModel}, which describes the balance of forces for fiber $i$, can be written as
\begin{align}
 (k_i + \T k_i) \xi_i + \sum_{j \in N_i} \eij \Sij [\xi_i - \xi_j] = \T k_i u,
 \label{e:fiber}
\end{align}
where $N_i$ denotes the set of indices $j$ for which $a_{ij}>0$ (i.e., $N_i$ is the set of fibers interacting with fiber $i$ or, using different terminology, the set of neighbors of node $i$ in the network with the adjacency matrix $\eij$ where each fiber is represented by a node).

Equation~\eqref{e:fiber} represents a piecewise linear system, which we can solve in each of the linear regimes while tracking the transitions from one linear regime to another. A switch between linear regimes occurs when any of the stop operators $\Sij$ saturates (i.e., when the magnitude of the friction force between any pair of fibers $i$ and $j$ achieves its maximal possible value $r_{ij}$) or desaturates (the magnitude of the friction force becomes smaller than $r_{ij}$); we describe this by saying that link $ij$ saturates or desaturates. Before we consider the transitions between linear regimes in more detail, let us write Eq.~\eqref{e:fiber} in the form of a linear matrix equation
\begin{align}
M \xbar = \T K u + \Dbar,
\label{e:matrix_form}
\end{align}
where $\xbar = \{\xi_1,\ldots, \xi_n\}$ and $\T K = \{ \T k_1, \ldots, \T k_n \}$. The matrix $M$ and vector $\Dbar$ take specific values [given by Eqs.~\eqref{e:elements_M} and~\eqref{e:elements_D} below] for each of the linear regimes.

We introduce a new quantity $\Oij$ which denotes the current reference point (the origin) for the interaction $\Sij[\xi_i-\xi_j]$ between nodes $i$ and $j$. Specifically, $\Oij$ is the value of $\XXij$ at which $\Sij[\xi_i-\xi_j]=0$, provided that the relative displacement $\XXij$ approaches the value $\Oij$ monotonically from its current value. Notice that $\Oij = - \Oji$. We also introduce a binary quantity $\lij$ to represent the current state of link $ij$ (interaction between fibers $i$ and $j$),
\begin{align}
 \lij =\lb\{\begin{array}{cl} 1\,,& \text{if link $ij$ is unsaturated} \\
                              0\,,& \text{if link $ij$ is saturated} \end{array}\rb. \,.
\end{align}
We assume that initially $\Oij=0$ for all the links and $\lij=1$ (all links are unsaturated). These quantities will be updated according to the rules described below when the variations in the input parameter $u$ become sufficiently large.

If a link $ij$ is unsaturated ($\lij=1$), then the value of $\Sij$ is given by $(\XXij-\Oij)$. In the case when link $ij$ is saturated ($\lij=0$), the value of $\Sij$ is given by $\rij\, \sgn(\XXij-\Oij)$. Therefore, using the notation $\Oij$ and $\lij$, we can rewrite Eq.~\eqref{e:fiber} as
\begin{align} \label{e:fiber_lin}
(k_i + \T k_i) \xi_i + &\sum_{j \in N_i} \lij \eij (\xi_i - \xi_j - \Oij)+ \\
\nn &\sum_{j \in N_i} (1-\lij) \eij \rij\, \sgn(\XXij- \Oij) = \T k_i u.
\end{align}
Equation~\eqref{e:fiber_lin} can be written in matrix form~\eqref{e:matrix_form} where the elements of $M$ and $\Dbar$ are given by
\begin{align}
\label{e:elements_M}
 M_{ij} =\lb\{\begin{array}{cl} -\eij \lij\,,& \text{if $i \ne j$ } \\
              k_i + \T k_i + \sum_{j \in N_i} \eij \lij\,,& \text{if $i=j$} \end{array}\rb. \,.
\end{align}
and
\begin{align}
\label{e:elements_D}
 \Dbar_i = \sum_{j \in N_i} \eij \lb(\lij \Oij - (1-\lij)\rij \sgn(\XXij- \Oij)\rb).
\end{align}
For example, if we consider three fibers connected as in Fig.~\ref{fig1}(b), then Eq.~\eqref{e:matrix_form} takes the form
\begin{widetext}
 \begin{align}
\lb(
\begin{array}{c c c}
\nn k_1 + \T k_1 + \sum\limits_{j \in N_1}\eoj\loj & -a_{12}l_{12} & -a_{13}l_{13}\\
-a_{21}l_{21} & k_2 + \T k_2 + {\sum \limits_{j \in N_2}} \edj\ldj& -a_{23}l_{23}\\
-a_{31}l_{31} & -a_{32}l_{32} & k_3 + \T k_3 + \sum\limits_{j \in N_3}\etj\ltj\\
\end{array}
\rb)
\lb(
\begin{array}{c}
 \xi_1\\
 \xi_2\\
 \xi_3
\end{array}
\rb)
= \\
\lb(
\begin{array}{c}
\T k_1\\
\T k_2\\
\T k_3
\end{array}
\rb)u+
\lb(
\begin{array}{c}
\sum\limits_{j \in N_1}\eoj \lb[l_{1j} \Ooj - (1-l_{1j}) r_{1j} \sgn(\xi_1-\xi_j- \Ooj) \rb]\\
\sum\limits_{j \in N_2}\edj \lb[l_{2j} \Odj - (1-l_{2j}) r_{2j} \sgn(\xi_2-\xi_j- \Odj) \rb]\\
\sum\limits_{j \in N_3}\etj \lb[l_{3j} \Otj - (1-l_{3j}) r_{3j} \sgn(\xi_3-\xi_j- \Otj) \rb]
\end{array}
\rb).
\label{main_Eq3}
\end{align}
\end{widetext}

Suppose we want to calculate the values of $\xi_i$ as the input $u$ varies. The solution of Eq.~\eqref{e:matrix_form} is given by
\begin{align} \label{e:xbar}
 \xbar &= M^{-1} (\T K u + \Dbar ).
\end{align}
However, we need to update $M$ and $\Dbar$ each time a link saturates or desaturates.

The condition for the saturation of an unsaturated link $ij$ is $\XXij = \Oij \pm \rij$. We note that when we check this condition for all pairs of $i$ and $j$, then it is sufficient to consider only one of the two cases, for example,
\begin{align} \label{e:sat_cond}
\XXij = \Oij + \rij,
\end{align}
since the other case is captured due to $\Oij - \rij = -(\Oji + \rji)$. Using the link saturation condition~\eqref{e:sat_cond} and Eq.~\eqref{e:xbar} we obtain the values of $\uij$ at which the link between nodes $i$ and $j$ saturates:
\begin{align}
 \uij = \frac{\Oij + \rij + (M^{-1} \Dbar)_i - (M^{-1} \Dbar)_j} {(M^{-1} \T K)_i - (M^{-1} \T K)_j}.
\end{align}
Hence, we can calculate $\xi_i$ from Eq.~\eqref{e:xbar} for all $u$ (without the need to update $M$ and $\Dbar$) until $u$ passes through any of $\uij$ values. When $u$ reaches any of $\uij$, this will indicate that we transition to a new linear regime and thus have to calculate new $M$ and $\Dbar$ as $l_{ij}$ changes from $1$ to $0$ at this point.

\setlength{\tabcolsep}{0.2cm}
\begin{table}[floatfix]
\begin{center}
\begin{tabular}{r|l r | l}
Step & $u$ & $\xi_1$ & Comments \\
\hline
0 &  0 & 0 & $u$ starts decreasing\\
1 & -1.56 & -0.5 & $S_{r_{23}}$ saturates to +$r_{23}$\\
2 & -2.02 & -0.65 & $S_{r_{13}}$ saturates to +$r_{13}$\\
3 & -33 & -2 & $S_{r_{12}}$ saturates to +$r_{12}$\\
4 & -100 & -2 & $u$ changes direction,\\
&&& $S_{r_{12}}$ remains saturated\\
5 & -96.97 & -0.77 & $S_{r_{23}}$ saturates to -$r_{23}$\\
6 & -95.2 & 0 & $S_{r_{13}}$ saturates to -$r_{13}$ causing\\
&&&subsequent desaturation of $S_{r_{12}}$\\
7 & -80 & 0.66 & $u$ changes direction\\
8 & -83.11 & -0.35 & $S_{r_{23}}$ saturates to +$r_{23}$\\
9 & -84.04 & -0.65 & $S_{r_{13}}$ saturates to +$r_{13}$\\
10 & -100 & -1.34 & $u$ changes direction\\
11 & -96.89 & -0.33 & $S_{r_{23}}$ saturates to -$r_{23}$\\
12 & -96.12 & -0.08 & $S_{r_{12}}$ saturates to +$r_{12}$\\
13 & -95.93 & 0 & $S_{r_{13}}$ saturates to -$r_{13}$ causing\\
&&&subsequent desaturation of $S_{r_{12}}$\\
14 & -90 & 0.26 & $u$ changes direction\\
15 & -93.11 & -0.75 & $S_{r_{23}}$ saturates to +$r_{23}$\\
16 & -94.04 & -1.05 & $S_{r_{13}}$ saturates to +$r_{13}$\\
17 & -97 & -1.18 & $u$ changes direction\\
18 & -93.89 & -0.17 & $S_{r_{23}}$ saturates to -$r_{23}$\\
19 & -92.96 & 0.13 & $S_{r_{13}}$ saturates to -$r_{13}$\\
20 & -75 & 0.91 & end of simulation
\end{tabular}
\end{center}
\caption{Table presenting the sequence of input values $u$, and the corresponding $\xi_1$ values, at which stop operators $S_{r_{ij}}$ saturate or desaturate for the example shown in the upper panel of Fig.~\ref{f:mech_non_PI}. Each saturation or desaturation creates a corner point of the piecewise linear trajectory. Steps 4 through 10 correspond to the non-closed loop shown by the bold line. The sequence of saturations for the lower panel of the same figure is as follows: $S_{r_{23}}$ (at $u\approx-1.56$), $S_{r_{13}}$ (at $u\approx-2.02$), $S_{r_{12}}$ (at $u=-33$). }
\label{table1}
\end{table}

The desaturation of a saturated link $ij$ occurs when $\XXij$ has a turning point (passes through a local maximum or minimum value). There are two ways this can happen. First, due to complex interactions between the nodes, a link $ij$ may desaturate due to the saturation of another link $mn$ (this happens in the example shown in Fig.~\ref{f:mech_non_PI} as described in steps 6 and 13 of Table~\ref{table1}). Second, saturated links may desaturate when the input $u$ has a turning point. (Interestingly, saturated links may remain saturated when $u$ makes a turning point; this happens in the example shown in Fig.~\ref{f:mech_non_PI} as described in step 4 of Table~\ref{table1} where $S_{r_{12}}$ does not desaturate.) In both cases, we need to determine whether $\XXij$ has a turning point by evaluating the sign of the derivative of $\XXij$ with respect to $u$. The derivative is obtained from Eq.~\eqref{e:xbar} and is given by
\begin{align}
 \XXderF = (M^{-1}\T K)_i - (M^{-1}\T K)_j.
\end{align}

In the first case, we need to evaluate the sign of $\XXderF$ before and after the saturation of $mn$. Moreover, a change in $\lij$ will affect matrix $M$ and therefore further changes in $\XXderF$ (and thus in $\lij$) are possible. This means that we need to iterate the evaluation of $\XXderF$, $\lij$ and $M$ until $\lij$ reaches a steady state.

In the second case, we need to find a partition of previously saturated links into a set of links that remain saturated and a set that becomes desaturated. These sets should ensure the consistency condition on $\XXderF$ when $u$ makes a turning point that $\XXderF$ should change the sign for links that remain saturated and not for links that become desaturated. Similarly to the first case, finding the set of desaturating links may be not straightforward because of the dependency of $\XXderF$ on $\lij$. However, this can be done numerically by simply looping through all possible partitions and finding the one that leads to consistency.

Finally, for the resulting set of links that became desaturated we calculate the new $\Oij$ from $\XXij$, $\rij$, and the current $\Oij$ as follows:
\begin{align}
\Oij^{\rm new} = \XXij - \sgn(\XXij-\Oij)\rij.
\end{align}

The above algorithm has been used to produce Figs.~\ref{f:mech_non_PI} and~\ref{f:mech_PI}. For example,
Table~\ref{table1} presents the sequence of input values $u$ at which stop operators $S_{r_{ij}}$ saturate or desaturate for the example shown in Fig.~\ref{f:mech_non_PI}.

\section{Analysis of the pricing model} \label{appC}
In this section we discuss in more detail the pricing model $r^*(t)=r(t)+\kappa\sigma(t)$, where $r^*$ is the log-price of the asset, $r$ is the exogenous Brownian information stream, the parameter $\kappa$ quantifies the effect of momentum traders on the price, and the sentiment of the market $\sigma$ is defined as the arithmetic mean of the states $\chi_i$ of momentum traders,
\begin{equation}\label{pd}
\sigma=\frac1N\sum_i^N \chi_i(t).
\end{equation}
Dynamics of the states are driven by the log-price according to the PI input-output relationship, $\chi_i(t)=I_{H_i}[r^*](t)$, which closes the model. Here the PR function $H_i(r^*)=H(r^*-\rho_i)$ is the step function with threshold $\rho_i$ chosen uniformly from $[c,a]$.

Testing the system with an increasing input, we see that in the continuum limit $N\to\infty$ the exogenous Brownian input and the variables $\sigma$ and $r^*$ are related by the formulas
\begin{align}
\sigma(t)=I_{\hat R} [r+\kappa \sigma](t),\quad r^*(t)=r(t)+\kappa I_{\hat R}[r^*](t)
\end{align}
where the PR function of the PI operator $I_{\hat R}$ has the profile shown in Fig.~\ref{f_PrimResp}(d) with $\rho_1=c$ and $\rho_2=a$. According to our results, these relationships can be easily solved explicitly,
\begin{equation}\label{pd1}
\sigma(t)=I_{R} [r](t),\quad r^*(t)=r(t)+\kappa \sigma(t)
\end{equation}
and two cases are possible. In the subcritical case, $\kappa<\kappa_c=(a-c)/2$, the PR function $R$ in these relationships also has the shape shown in Fig.~\ref{f_PrimResp}(d) with the same $\rho_1=c$, but with a smaller $\rho_2=a-2\kappa>\rho_1$. In the supecritical case $\kappa>\kappa_c$, the function $R$ is the step function with the threshold $c$. That is, in the supercritical case, due to a global avalanche, all the traders switch their state simultaneously causing $\sigma$ to jump between the values $\pm1$. The statistics of the intervals between jumps can be found by solving an exit time problem.
\begin{figure}[!t]
\includegraphics[width=0.7\columnwidth]{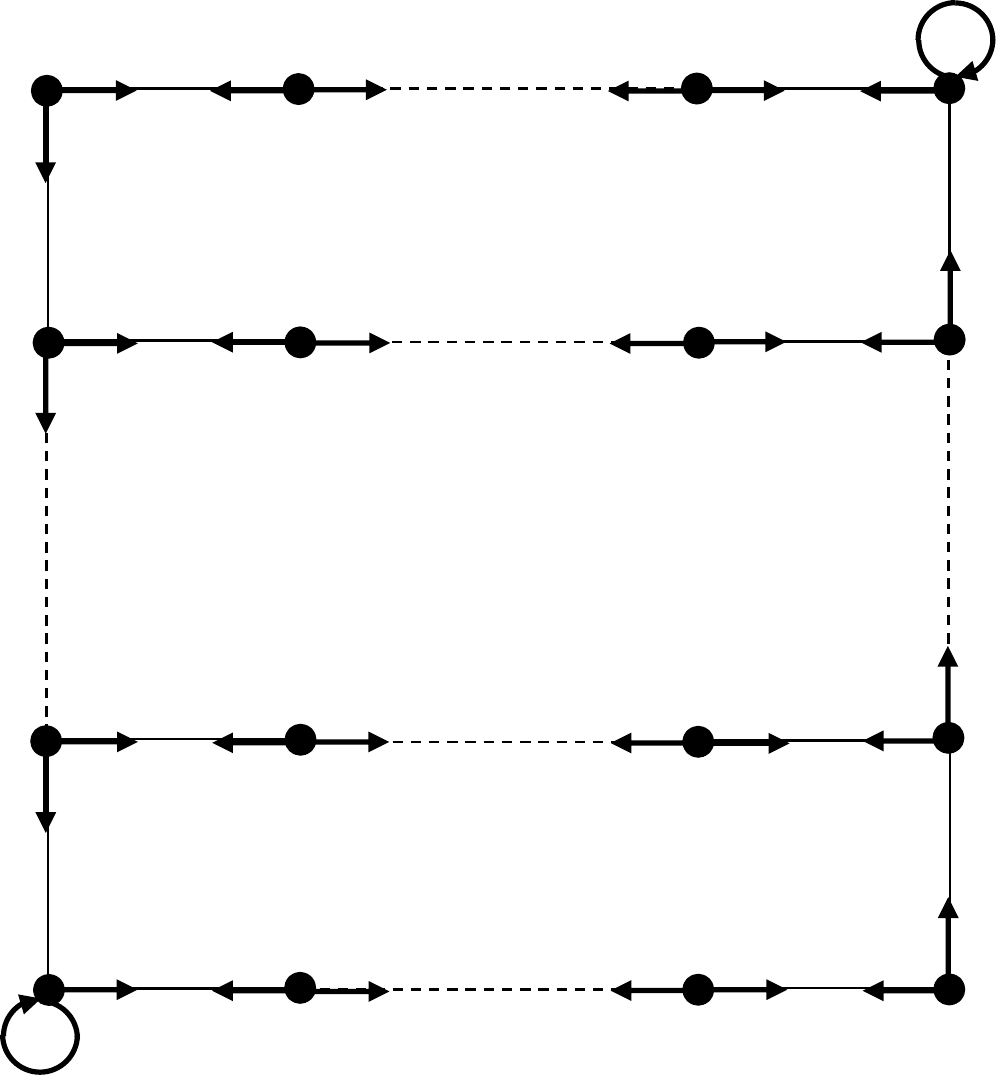}
\caption{Rectangular random walk $(w,\sigma)$. At each time step, a particle makes one of two possible moves with equal probability as shown by arrows. It moves either to a neighboring node or, if it is at the upper right or lower left corner, possibly to the same node.}
\label{rectangle}
\end{figure}

We first consider the subcritical case which is more relevant to normal market conditions and more interesting. Our objective is to calculate the profile of the daily log-price increments histogram shown in Fig.~\ref{f:pricing_model}(c). For this purpose, we first find the stationary distribution of the stochastic process $\sigma(t)=I_{R} [r](t)$. The shape of the PR curve $R$ allows us to describe this process as a random walk of a particle in a rectangle, where the vertical coordinate of the particle is $\sigma$, while the horizontal coordinate is an auxiliary variable $w$, see Fig.~\ref{rectangle}. The motion of the particle $(w(t),\sigma(t))$ is driven by the Brownian input $r(t)$. For simplicity, we describe the random walk in a discrete time and state setting. In this case, the particle lives on a rectangular mesh with $n_x$ columns and $n_y$ rows and the Brownian input is represented by a random walk $r$ which at every time step with equal probability makes one step left or one step right along a uniform mesh on the real line. First assume that the input $r$ moves left at some moment. Then, if the particle was not on the left side of the rectangle (left column of the mesh), it also moves one step left to a neighboring node; it moves one step down from any node of the left side, except from the lower corner; and, if the particle was in the lower left corner of the rectangle, it remains there. Similarly, when $r$ moves right, so does the particle if it was not on the right side of the rectangle; it moves one step up from any node of the right side, except from the upper corner; and it remains in the upper right corner if it was there (see Fig.~\ref{rectangle}). In this model, the horizontal and vertical step of the rectangular mesh are related by $|\Delta w|=(\kappa_c-\kappa)|\Delta\sigma|$, the horizontal step equals the step of the input mesh, $|\Delta w|=|\Delta r|$, and the number of rows and columns in the rectangular mesh are related by $c n_y = 2(\kappa_c-\kappa) n_x$. These relationships ensure that the increment of the log-price equals $\Delta r^*=\Delta r+ \kappa\Delta\sigma$, where $\Delta r$ and $\Delta \sigma$ are the increments of the input and the vertical coordinate of the particle at the same time step, respectively.

A simple calculation shows that the probability density of the stationary distribution for the random walk $(w,\sigma)$ linearly decreases on the lower (upper) side of the rectangle from the lower left to the lower right (upper right to upper left) corner and is uniform on the rest of the rectangle. In the continuous time and state limit ($n_x,n_y\to\infty$), when the input $r(t)$ becomes the continuous Brownian motion, the density function of the stationary probability distribution for the random process $(w(t),\sigma(t))$ on the rectangle $\Pi=\{0\le w\le c,\ 0\le\sigma\le 2\}$ is \begin{equation}\label{rhost}
\rho_{st}(w, \sigma)=\frac{(c-w)\delta(\sigma)+w \delta(\sigma-2) +\kappa_c-\kappa}{c(a-2\kappa)},
\end{equation}
where $\delta$ denotes the Dirac $\delta$ function. We note that in the continuum limit the process $w$ becomes the reflected Brownian motion on the interval $[0,c]$ (with reflecting boundary condition at both ends).

Calculations of the profile of the histogram for daily log-price increments $\Delta r^*_n=r^*(t_n+\tau)-r^*(t_n)$, where $\tau=1$\,day is a fixed time interval and $t_n=n\tau$, will be performed in the continuous time and state setting. Assuming ergodicity, statistics of the increments $\Delta r_n$ obtained from a typical long trajectory of the processes $r$, and $(w,\sigma)$ can be approximated by the probability density function of the random variable
\begin{equation}\label{incre}
\Delta r^* = r^*(\tau)-r^*(0)=r(\tau)+\kappa (\sigma(\tau)-\sigma(0)),
\end{equation}
where the stationary process $(w(t),\sigma(t))$ bounded by the rectangle $\Pi$ is driven by the Brownian input $r(t)$ [with $r(0)=0$] and has the law~\eqref{rhost}. The following calculations are based on the assumption that the maximal increment of the Brownian input $r$ during 1 day remains bounded by the quantity $c/2$ with a probability close to 1,
\begin{equation}\label{ll}
P\lb(\max_{0\le t\le \tau}|r(t)|\ge c/2\rb)\ll 1.
\end{equation}
For the plots shown in Fig.~\ref{f:pricing_model}, the variance of the Brownian input $r(T)$ at the end of the time interval $T=40$ years (with 250 trading days per year) has been set to 1. Hence, for one trading day $r(\tau)\sim N(0, \Sigma^2)$ with the standard deviation $\Sigma=0.01$. Since $c/2=2.5\Sigma$ for these plots, $P( |r(\tau)|\le c/2)=0.988$, which agrees with~\eqref{ll}. We will consider only those input trajectories that satisfy $|r(t)|< c/2$ on the whole time interval $0\le t\le \tau$. The corresponding trajectories of the process $(w,\sigma)$ cannot reach both left and right sides of the rectangle $\Pi$ during the same time interval. Trajectories for which this occurs will be disregarded.

Thus, let us consider trajectories $(w(t),\sigma(t))$ corresponding to different realizations of the Brownian $r(t)$ on the time interval $0\le t\le \tau$ and different initial data $(w(0),\sigma(0))$, restricting our attention to initial data from the right half of the rectangle $\Pi$, i.e., with $c/2\le w(0)\le c$, $0\le \sigma(0)\le 2$. (Trajectories starting at the left half of $\Pi$ can be treated similarly). Since we assume that $r(t)>- c/2$ for all $0 \le t \le \tau$ [other inputs are disregarded due to~\eqref{ll}], a trajectory starting from the right half of $\Pi$ never reaches the left side of the rectangle. For such trajectories, the log-price increment~\eqref{incre} can be easily expressed in terms of the variables $w(0)$, $\sigma(0)$, $r(\tau)$, and $m(\tau)=\max_{0\le t\le\tau} r(t)$, the maximum input value, where the probability density of the joint distribution for the Brownian motion and its running maximum is defined by the relation
\begin{align}
\rho_{br}(r,m)=\lb\{
\begin{array}{ll}\frac{2(2m-r)}{\tau\sqrt{2\pi \tau}}e^{-\frac{(2m-r)^2}{2\tau}},& m\ge 0, m\ge r,\\
0,& {\rm otherwise}.
\end{array}\rb.
\end{align}
The two-dimensional random variable $(r(\tau), m(\tau))$ and the two-dimensional variable $(w(0),\sigma(0))$, which has the law~\eqref{rhost}, are independent. As the expression for $\Delta r^*$ depends on relations between these variables, we classify trajectories into a few groups.

If $\sigma(0)=2$, then the trajectory remains on the upper boundary of the rectangle $\Pi$ all the time ($\sigma(t)=2$ for all $0\le t\le\tau$), hence the log-price increment~\eqref{incre} equals the increment of the input, $\Delta r^*=r(\tau)$. Since $r(\tau)$ is normally distributed, so is $\Delta r^*$ for such trajectories,
\begin{equation}\label{case1}
\rho(\Delta r^*=y, \sigma(0)=2)=\frac{3c}{8(a-2\kappa)}\cdot \frac{e^{-\frac{y^2}{2\tau}}}{\sqrt{2\pi \tau}},
\end{equation}
where $P=3c/(8(a-2\kappa))$ is the total probability to find the point $(w(0),\sigma(0))$ on the right half of the upper side of the rectangle, see~\eqref{rhost}.

Another class consists of trajectories that start below the upper side of the rectangle $\Pi$ and never reach its right side during the day. This class is defined by the relations
\begin{equation}\label{case2defi}
0\le \sigma(0)<2; \quad c/2\le w(0)<c-m(\tau).
\end{equation}
For such trajectories, $\sigma(t)=\sigma(0)$ for all $0\le t\le\tau$ and hence $\Delta r^*=r(\tau)$, as in the previous case. Integrating the product of the probability densities $\rho_{st}(w,\sigma)\rho_{br}(r,m)$ over domain~\eqref{case2defi} with respect to the variables $w(0)=w$, $\sigma(0)=\sigma$, and $m(\tau)=m$, we obtain the probability density function of the log-price increment for this class of trajectories. After some manipulations,
this probability density can be presented as the integral
\begin{widetext}
\begin{equation}\label{sr}
\rho(\Delta r^*=y, \sigma(0)<2, w(0)<c-m(\tau))=\frac{1}{2c(a-2\kappa)}
\int_{\max\{0,y\}}^{c/2} \rho_{br}(y,m)\lb(\frac{c}2-m\rb)\lb(\frac{c}2+m+4(\kappa_c -\kappa)\rb)\,dm,
\end{equation}
\end{widetext}
which can be expressed explicitly in terms of the Gaussian and the error function.

The next set of conditions,
\begin{equation}\label{mc}
m(\tau)+w(0)>c; \quad \frac{m(\tau)+w(0)-c}{\kappa_c-\kappa}<2-\sigma(0),
\end{equation}
ensures that a trajectory reaches the right side but not the upper side of the rectangle $\Pi$. For such trajectories,
\begin{equation}\label{po}
 \Delta r^*=r (\tau) + \frac{\kappa}{\kappa_c-\kappa} (m(\tau)+w(0)-c).
\end{equation}
Hence, we obtain the probability density function $\rho(\Delta r^*=y)$ of the log-price increment for this class by integrating the product $\rho_{st}(w,\sigma)\rho_{br}(r,m)$, where $r=r(\tau)$ is related to the variables $w=w(0), \sigma=\sigma(0), m=m(\tau)$ by formula~\eqref{po} with $\Delta r^*=y$ kept fixed; relations~\eqref{mc} define the domain of integration in the product of the domain $\Pi$ of the pair $(w,\sigma)$ and the line $m$. The resulting triple integral can be reduced to the sum of the following two terms:
\begin{widetext}
\begin{equation}\label{sr1}
\rho(\Delta r^*=y, 0<\sigma(0)<\sigma(\tau)<2)=c_0\int_{\max\{0, (y-c/2)/\kappa_c\}}^2 (2-p)\, dp \int_{\max\{0, y-\kappa_c p\}}^{c/2} \rho_{br}\bigl(y-\kappa p, q+(\kappa_c-\kappa)p\bigr)\,dq,
\end{equation}
\begin{equation}\label{sr2}
\rho(\Delta r^*=y, 0=\sigma(0)<\sigma(\tau)<2)=\frac{c_0}{\kappa_c-\kappa}\int_{\max\{0, (y-c/2)/\kappa_c\}}^2 dp \int_{\max\{0, y-\kappa_c p\}}^{c/2} q\, \rho_{br}\bigl(y-\kappa p, q+(\kappa_c-\kappa)p\bigr)\,dq,
\end{equation}
\end{widetext}
where $c_0=(\kappa_c-\kappa)^2/(c(a-2\kappa))$.

Finally, there are trajectories starting below the upper side of the rectangle that reach the upper side during the day. This class is defined by the conditions
\begin{equation}
0< 2-\sigma(0)<\frac{m(\tau)+w(0)-c}{\kappa_c-\kappa}
\end{equation}
and the corresponding log-price increment equals $\Delta r^*=\Delta r+ \kappa (2-\sigma(0))$. For the subcritical parameter set we consider, the probability of having such trajectories is small and their contribution has almost no effect on the profile of the probability density plot. Hence, we have discarded a correction to the probability density function of $\Delta r^*$ due to such trajectories.

Thus, denoting the sum of contributions~\eqref{case1}, \eqref{sr}, \eqref{sr1}, and \eqref{sr2} from different classes of trajectories starting in the right half of $\Pi$ by $\rho_r(y)$, the symmetrized sum
\begin{equation}\label{symm}
\rho(\Delta r^*=y)=\rho_r(y)+\rho_r(-y)
\end{equation}
provides an analytic approximation to the probability density function of the log-price daily increments, see the theoretical curve in Fig.~\ref{f:pricing_model}(c). The term $\rho_r(-y)$ accounts for trajectories starting in the left half of $\Pi$.

We now look at the critical value $\kappa=\kappa_c$. In the critical case, each trajectory that reaches the right side of the rectangle immediately jumps to its upper side. Hence, $\rho_r(y)$ is the sum of expressions~\eqref{case1} and~\eqref{sr} only [with no terms of the form~\eqref{sr1} and~\eqref{sr2}]. The symmetrized sum~\eqref{symm} describes the main central mode of the probability density distribution shown in Fig.~\ref{f:pricing_model}(d). One small side mode appears due to trajectories that start on the lower side of the rectangle and reach (jump to) the upper side, that is, trajectories that have been disregarded in the subcritical case. The profile of the side modes is described by the left and right shifts $\rho_{side}(\pm(y+2\kappa))$ of the function
\begin{align}
\rho_{side}(y)=\frac1{2c^2}\int_{\max\{y,0\}}^{c/2} m^2\rho_{br}(y,m)\,dm.
\end{align}
Hence, the central mode and side modes can be explicitly expressed as a combination of the Gaussian and the error function.

In the supercritical case $\kappa>\kappa_c$, the central mode is the same as in the critical case, while the side modes have the same shape as in the critical case but shift further to left and right.

\bibliography{networks}

\end{document}